\documentclass[sigconf]{acmart}

\usepackage[T1]{fontenc}
\usepackage{booktabs} 
\usepackage{romannum}
\usepackage{subcaption}
\usepackage{balance}

\copyrightyear{2017} 
\acmYear{2017} 
\setcopyright{acmcopyright}
\acmConference{MM '17}{October 23--27, 2017}{Mountain View, CA, USA}\acmPrice{15.00}\acmDOI{10.1145/3123266.3123429}
\acmISBN{978-1-4503-4906-2/17/10}

\begin{document}
\title{Cross-Domain Image Retrieval \\ with Attention Modeling}

\author{Xin Ji, Wei Wang}
\affiliation{%
	\institution{National University of Singapore
	}
}
\email{{jixin, wangwei}@comp.nus.edu.sg}

\author{Meihui Zhang}
\authornote{Meihui Zhang is the corresponding author.}
\affiliation{
	\institution{Singapore University of Technology and Design}
}
\email{meihui\_zhang@sutd.edu.sg}

\author{Yang Yang}
\affiliation{
	\institution{University of Electronic Science and Technology of China}
}
\email{dlyyang@gmail.com}

\begin{abstract}
With the proliferation of e-commerce websites and the ubiquitousness of smart phones, cross-domain image retrieval using images taken by smart phones as queries to search products on e-commerce websites is emerging as a popular application. One challenge of this task is to locate the attention of both the query and database images. In particular, database images, e.g. of fashion products, on e-commerce websites are typically displayed with other accessories, and the images taken by users contain noisy background and large variations in orientation and lighting. Consequently, their attention is difficult to locate. In this paper, we exploit the rich tag information available on the e-commerce websites to locate the attention of database images. For query images, we use each candidate image in the database as the context to locate the query attention. Novel deep convolutional neural network architectures, namely TagYNet and CtxYNet, are proposed to learn the attention weights and then extract effective representations of the images. Experimental results on public datasets confirm that our approaches have significant improvement over the existing methods in terms of the retrieval accuracy and efficiency. 
\end{abstract}

%
%




\maketitle

\section{Introduction}

With the ubiquitousness of smart phones, it is convenient for people to take photos anytime and anywhere. For example, we usually take photos of beautiful sceneries in our outings, candid photos of funny moments, and well-presented dishes in restaurants. 
These photos can be used as queries to search visually similar images on the Internet. 
With the wide acceptance of e-commerce, product image search becomes part-and-parcel of online shopping, where users submit a photo taken via their cell phones to look for visually similar products~\cite{huang2015cross}.

\begin{figure}
	\centering
	\includegraphics[width=0.4\textwidth]{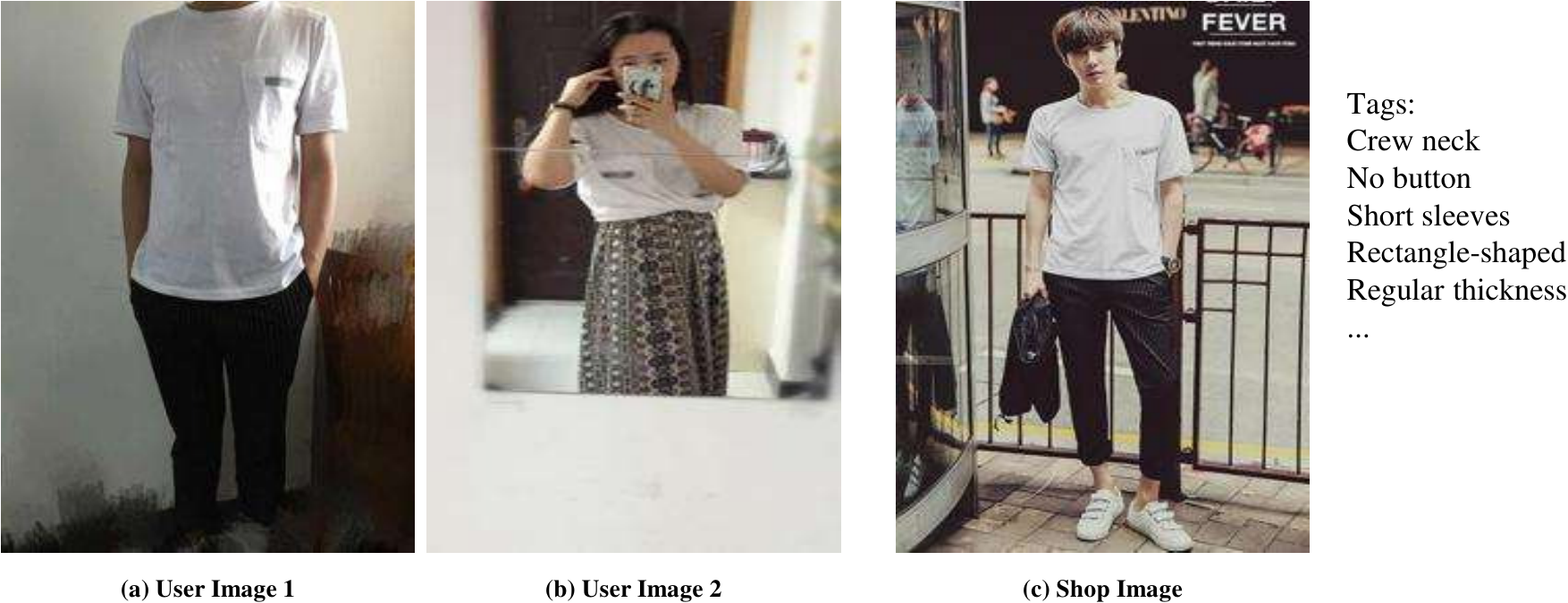}
	\caption{Example images of the same product.}\label{fig:example}
\end{figure}

Product image search is a challenging problem as it involves images from two heterogeneous domains, namely the user domain and the shop domain. The user domain consists of the query images taken by users, and the shop domain consists of the database images taken by professional photographers for the e-commerce websites. Images from the two domains exhibit different characteristics. For instance, images from user domain (see Figure~\ref{fig:example}(a), \ref{fig:example}(b)) typically have large variations in orientation and lighting, whereas images from shop domain are taken under good condition(see Figure~\ref{fig:example}(c)). How to model the domain-specific and domain-invariant features is a challenge. Traditional image retrieval approaches designed for a single domain cannot model the domain-specific features of different domains. DARN\cite{huang2015cross} and MSAE~\cite{wang2014effective,wang2016effective} create different deep learning branches for each domain separately to model domain-specific features. However, they cannot capture the common features shared by both domains.

Another challenge comes from the difficulty of attention modeling for the user domain and the shop domain. Attention modeling is to find the focus of images to extract effective image representations from the salient areas, i.e. spatial locations. For images from user domain, however, they exhibit large variations and have noisy background as shown by the example in Figure~\ref{fig:example}(a) and \ref{fig:example}(b). 
For shop images, it is common that the whole-body clothes along with some accessories are displayed in an image as shown in Figure~\ref{fig:example}(c), whereas the target is the upper clothes.
Without external information, it is difficult to find the attention of the query image (from the user domain) and the database image (from the shop domain). Existing approaches, including DARN~\cite{huang2015cross} and FashionNet~\cite{liu2016deepfashion}, reply on human annotated bounding boxes and landmarks for attention modeling. However, it is costly to do such annotations for large image databases from e-commerce websites.

In this paper, we propose a neural network architecture to extract effective image representations for cross-domain product image search. Inspired by \cite{yosinski2014transferable}, which shows that the bottom layers of a convolutional neural network (CNN) learn domain-invariant features and the top layers learn domain-specific features, we propose a `Y-shape' network architecture that shares a set of convolution layers at the bottom for both domains, and has separate branches on the top for each domain. Each branch on the top contains an attention modeling layer to learn the spatial attention weights of the feature maps from the convolution layer. The attention modeling layer of the shop branch exploits the tag information of shop images, which are widely available on e-commerce websites. Tags like product category and attributes can help us to locate the attention easily. For instance, in Figure~\ref{fig:example}(c), from the tag `Flare Sleeves', we know that the focus of this image is sleeves rather than trousers. DARN and FashionNet also use tags to train their models. Different from them, we use tags as inputs instead of prediction outputs. 
For the attention modeling layer of the user branch, since user-provided tags are not widely available and usually contain much noise, we use the candidate shop image as the context to help locate the attention. For example, in Figure~\ref{fig:example}(b), the focus of the user image is not clear (T-shirt or dress) until we compare it with the shop image. The final image representation is generated by aggregating the features at each spatial location according to the attention weight.

We adapt the triplet loss~\cite{schroff2015facenet} as the training objective function. A training triple consists of an anchor image from the user domain, a positive image from the shop domain that has the same product as the anchor image, and a negative image from the shop domain that does not have the same product as the anchor image. We first forward them through the shared layers to get a set of feature maps for each image. Second, we forward the feature maps of the positive (resp. negative) image and its tags into the shop branch. Third, the generated representation of the positive (resp. negative) image and the feature maps of the anchor image are fed together into the user branch to learn the representation of the anchor image w.r.t the positive (resp. negative) image. We extend the triplet loss to accept four inputs, namely the representations for the positive image, the negative image, and the anchor image w.r.t the positive and negative image. The training procedure updates the network parameters to move the similar pairs closer than dissimilar pairs. 

During querying, our approach needs to extract the query representation w.r.t each database image, which is expensive. To improve the search efficiency, we follow~\cite{cao2016focus} to do an initial search with the query representation generated by aggregating its feature maps equally across all spatial locations (i.e. all locations have the same attention weight~\cite{babenko2015aggregating}). The returned top-K candidate images are re-ranked using the query representation extracted by attention modeling.

Our contributions include:
\begin{enumerate}
	\item A novel neural network architecture to extract effective features for cross-domain product image retrieval. 
	\item Two attention modeling approaches for the user and shop domain respectively, i.e., exploiting tag information to help locate the attention of the database images (denoted as TagYNet) and exploiting the candidate database images to locate the attention of the query images (denoted as CtxYNet).
	\item Comprehensive experimental study, which confirms that our approaches outperform existing solutions\cite{huang2015cross,liu2016deepfashion} significantly.
\end{enumerate}

\section{Related Work}

Our cross-domain product image retrieval is one type of Content Based Image Retrieval (CBIR). CBIR has been studied for decades~\cite{datta2008image} to search for visually similar images, and it consists of two major steps. First, a feature vector is extracted to represent each image, including the query image and images in the database. 
Second, the similarity (e.g. cosine similarity) of the query image and the database images are computed based on the feature vectors. Top-K similar images are usually returned to the user. 
Many research works have been proposed to improve the search performance from the feature extraction and similarity measurement perspectives. For instance, local descriptors such as SIFT~\cite{lowe1999object}, and feature aggregation algorithm such as VLAD~\cite{jegou2010aggregating}, have been proposed for improving the feature representation. 
Metric learning~\cite{bellet2013survey} is a research area which focuses on learning mapping functions to project features into an embedding space for better similarity measurement. 
With the resurgence of deep learning~\cite{krizhevsky2012imagenet,wang2016database}, many approaches based on deep learning models have been proposed towards learning semantic-rich features and similarity metric. We review some related works as follows.

\subsection{Feature Learning}

Feature learning, in contrast to feature engineering, learns representation of data (i.e. images) directly. 
It becomes popular due to its superior performance over hand-crafted features, such as SIFT~\cite{lowe1999object}, 
for many computer vision tasks such as image classification~\cite{krizhevsky2012imagenet} and retrieval\cite{wan2014deep}.

\subsubsection{Deep Convolutional Neural Network}
Deep convolutional neural networks (DCNN) \cite{krizhevsky2012imagenet,he2016deep} consists of multiple convolutions, pooling, and other layers. These layers are tuned over a large training dataset to extract semantic-rich features for the tasks of interest, e.g. image classification. 
Rather than training a DCNN from scratch, recent works \cite{tang2016generalized} have shown that transfer learning, by introducing knowledge from large-scale well-labeled dataset such as ImageNet \cite{deng2009imagenet}, can achieve competitive performance with less training time.
Apart from transferring knowledge from domain to domain~\cite{yang2016zero,yang2014exploiting}, Ji et al.~\cite{wan2014deep} shows that introducing knowledge from one task to another, i.e., from image classification to image retrieval, also works well. 
For example, DARN~\cite{huang2015cross} and FashionNet~\cite{liu2016deepfashion} fine-tune the DCNNs to extract features for attribute prediction and cross-domain image retrieval. 
These work incorporate the attribute information into the model to capture more semantics in feature representation. 
In addition, FashionNet trains a subnetwork to predict the landmarks (a.k.a. attention areas) by extracting features from local areas. 
Different from these methods, we exploit the attributes of database images, which are easier to collect than landmarks, 
to locate the spatial attention areas of images directly and then extract features from these areas to construct our image feature for retrieval. 
Moreover, for the query image, we extract context-dependent attention when calculating its similarity with database images (see Section ~\ref{sec:user-attention}).

\subsubsection{Attention Modeling}
Attention modeling differentiates inputs with different weights. It has been exploited in computer vision to extract features from salient areas of an image~\cite{xu2015show}. 
Recently, it is applied in machine translation models~\cite{bahdanau2014neural} to locate the attention of source words for generating the next target word,
and in trajectory prediction to track the attention of dynamic objects like pedestrians and bikers~\cite{varshneya2017human}.
Typically, external information is required to infer the attention weights. 
For example, the image caption generation model~\cite{xu2015show} uses the previous word to infer the attention weights for features from different locations of the image, and then generates the next word using the feature from weighted aggregation. There are different approaches to compute the attention weights by aligning the external information with the visual feature of each location, 
e.g., via a multi-layer perceptron (MLP)~\cite{bahdanau2014neural}, inner-product~\cite{sukhbaatar2015end}, outer-product~\cite{fukui2016multimodal}, etc.
We explicitly exploit the image attributes as the external information to locate the attention of images in the database, and exploit the database image as the context to infer the attention of the query image.
Attention modeling ignores noisy background (occlusion), and thus extracts discriminative features for retrieval.

\subsection{Deep Metric Learning}\label{sec:metric}
Deep metric learning trains the neural networks to project the images into metric space for similarity measurement. 
Contrastive loss based on Siamese network \cite{chopra2005learning} is the most widely used pairwise loss for metric learning. Wang et al. \cite{wang2016matching} optimizes the contrastive loss by adding a constraint on the penalty to maintain robustness on noise positive pairs. Unlike pairwise loss that considers the absolute distance of pairs, triplet loss \cite{schroff2015facenet} computes the relative distance between a positive pair and a negative pair of the same anchor. Yuan et al. \cite{yuan2016hard} changes the original triplet loss into a cascaded way in order to mine hard examples. Song et al. \cite{oh2016deep} designs a mini-batch triplet loss that considered all possible triplet correlations inside a mini-batch. 
Liu et al. \cite{liu2016deep} proposes a cluster-level triplet loss that considered the correlation of cluster center, positive samples and the nearest negative sample. 
Query adaptive matching proposed in \cite{cao2016focus} computes the similarity based on the aggregated local features of each candidate image. The aggregation weights are determined by solving an optimization problem using the query as the context. However, it incurs extra computation time. 
In this paper, we generate different representations for the query image to compute its similarity with different database images. 
In other words, the query representation is adaptive to the compared database image.

\section{Problem Statement}
In this paper, we study the problem of cross-domain product image retrieval where the query image comes from user domain $I_u$ (i.e. taken by users) and the database images are from  shop domain $I_s$ (i.e. from the e-commerce websites). A database image is matched with the query image if they contain the same product. Our task is to train a model to extract effective image representations where matched pairs have smaller distance. 
We adopt CNN in our model for image feature extraction as CNN has shown outstanding performance in learning features for variant computer vision tasks~\cite{chatfield2014return}. Denote the feature vector at location $l$ of the convolution layer as $\mathbf{x}_l\in R^C$, where $C$ is the number of channels, i.e. the number of feature maps. To aggregate the features across all locations effectively, we train an attention modeling subnetwork for each domain to learn the spatial weights (i.e. attention). Once we get the weight of each location, denoted as $a_l\in R$, the final image feature is aggregated as $\sum_{l}a_l * \mathbf{x}_l$.

Our training dataset consists of user images and shop images, where each image has a product ID. The product ID is used for constructing triples for metric learning, where each triple consists of an anchor image $o\in I_u$, a positive image $p\in I_s$ which contains the same product as the anchor image, and a negative image $q\in I_s$ which does not contain the same product as the anchor image. In addition, each shop image $x$ has a set of tags (e.g. category or attributes) represented using a binary vector $\mathbf{x}^t\in \{0, 1\}^T$\footnote{with the value 1 indicting the presence of the tag and 0 otherwise}, where $T$ is the total number of tags. These tags are exploited for attention modeling (see Section~\ref{sec:shop-attention}). 

The notations used in this paper is summarized in Table~\ref{tb:notation}. 
Scalar values are in normal fonts, whereas bold fonts are for vectors and matrices. 
We refer to $o, p, q, x$ as images and $\mathbf{o,p,q,x}$ as image feature maps (i.e. output of convolution layers) or vectors. We reuse the notation $\mathbf{o,p,q,x}$ as the input and output of a subnetwork.

\begin{table}
	\centering
	\caption{Notations.}\label{tb:notation}
	\begin{tabular}{|l|l|}
		\hline
		$I_u$ & user domain images taken by users\\\hline 
		$I_s$ & shop domain images from e-commerce websites\\\hline 
		$o\in I_u$ & anchor image or query image \\\hline
		$p\in I_s$ & positive image (matched image to $o$) \\\hline
		$q\in I_s$ & negative image (unmatched image to $o$) \\\hline
		$\mathbf{o}^p$ (resp. $\mathbf{o}^q$) & feature of $o$ using the attention w.r.t to $p$ (resp. $q$)\\\hline
		$\mathbf{x}_l\in R^C$ & feature at location $l$, $x\in \{o, p, q\}$\\\hline
		$\mathbf{x}^t\in \{0, 1\}^T$ & tags vector, $x\in \{p, q\}$\\\hline 
		$T$ & total number of tags \\\hline 
		$a_l\in R$ & attention weight at location $l$ \\\hline 
		$\mathbf{W}\in R^{T*C}$ & tag embedding matrix\\\hline
		$\mathbf{v}\in R^{C}$ & weight vector in Equation~\ref{eq:gs} \\\hline
		$\mathbf{U}\in R^{L*C}$ & weight matrix in Equation~\ref{eq:gs} \\\hline
		$L$ & total number of locations, i.e. height*width \\\hline 
		$g_u$(), $g_s$() & feature alignment function in Equation~\ref{eq:gu} and \ref{eq:gs} \\\hline
	\end{tabular}
\end{table}


\section{Our Approach}\label{sec:approach}

In this section, we introduce our approach for inferring the spatial attention of both the query and database images to extract effective features for product image search. The network architecture, attention models, training and retrieval algorithms shall be discussed.

\subsection{Overview}

\begin{figure*}[htb]
	\centering
	\includegraphics[width=0.79\textwidth]{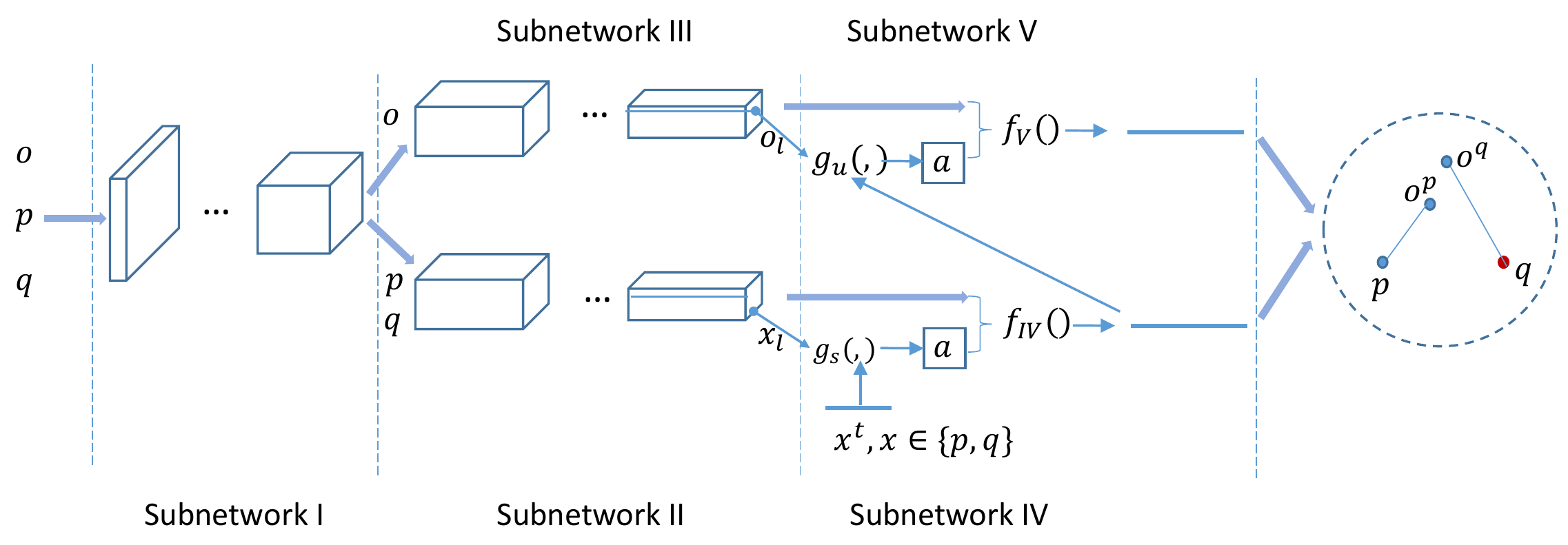}
	\caption{Illustration of our neural network architecture.}
   \label{fig:arch}
\end{figure*}

Our network architecture is illustrated in Figure~\ref{fig:arch}, which is like the character 'Y' in landscape. It consists of convolution layers (subnetwork \Romannum{1}, \Romannum{2} and \Romannum{3}) for image feature extraction, attention layers for spatial attention modeling (subnetwork \Romannum{4} and \Romannum{5}), and a triplet ranking loss layer for metric learning. 

During training, each training triple $<o, p, q>$ is forwarded through a shared subnetwork \Romannum{1} and then passed to two separate subnetwork \Romannum{2} and \Romannum{3}. This design is to learn domain-invariant features by \Romannum{1} and learn domain-specific features by subnetwork \Romannum{2} and \Romannum{3}. 
The shared subnetwork \Romannum{1} also saves memory compared with the dual network architecture in \cite{huang2015cross} which needs to store the parameters for both branches. We denote the subnetwork \Romannum{1} \Romannum{2} \Romannum{3} together as \textbf{YNet}. Subnetwork \Romannum{4} infers the spatial attention weights by exploiting the tags (e.g. attributes) of shop products. With the attention weights, it aggregates the features over all locations to get the feature vector of $p$  (resp. $q$). The subnetwork \Romannum{1} \Romannum{2} \Romannum{3} \Romannum{4} is denoted as \textbf{TagYNet}. By using $p$ (resp. $q$) as the context, subnetwork \Romannum{5} computes the attention of the anchor image $o$ w.r.t the $p$ (resp. $q$). The whole network is denoted as \textbf{CtxYNet}. The image representations are fed into an adapted triplet-loss function.

After training, we extract the representation of each database image via subnetwork \Romannum{1}, \Romannum{2} and \Romannum{4}. When a query arrives, we extract a simple representation of the query by forwarding the image through subnetwork \Romannum{1} and \Romannum{3}, and then averaging features at all spatial locations. This simple representation is used to conduct an initial retrieval. 
For each candidate image in the top-K (K=256 in our experiments) list, we use it to infer the query's attention via subnetwork \Romannum{5} and then compute the similarity. These images are finally re-ranked based on the new similarity score.

Specifications of the CNN layers in subnetwork \Romannum{1}, \Romannum{2} and \Romannum{3} will be given in Section~\ref{sec:exp}. We next introduce the subnetwork \Romannum{4}, \Romannum{5} for attention modeling and the adapted triplet loss function. 

\subsection{Attention Modeling for Shop Images}\label{sec:shop-attention}

Shop images are usually taken in good condition by professional photographers. Hence, they have better quality than user images. However, some images, especially for fashion products, are typically presented with accessories or other items as shown in Figure~\ref{fig:example}(c). To get the attention, we exploit the tags (or attributes) associated with the shop image, which can be collected easily from the shopping websites. Take the shop image in Figure~\ref{fig:example}(c) as an example, the tags `Crew neck', `Short sleeves' and `Rectangle-shaped' are useful for locating the attention of the image.

\begin{eqnarray}
g_s(\mathbf{x}_l, \mathbf{x}^t) &=& \mathbf{x}_l \cdot (\mathbf{W}^t \cdot \mathbf{x}^t) \label{eq:gu}\\
a_l &=& \frac{e^{g_s(\mathbf{x}_l, \mathbf{x}^t)}}{\sum_k^L e^{g_s(\mathbf{x}_k, \mathbf{x}^t)}} \label{eq:shop-attention}\\
f_{\Romannum{4}}(\mathbf{x}, \mathbf{a})&=&\sum_l^L a_l*\mathbf{x}_l \label{eq:shop-feature}
\end{eqnarray}

Given a shop image associated with the tag vector $\mathbf{x}^t\in \{0, 1\}^T$, we forward the image's raw feature (i.e. the RGB feature) through subnetwork \Romannum{1} and \Romannum{2} to get the feature maps $\mathbf{x}$. Subnetwork \Romannum{4} computes the attention of each spatial location through Equation~\ref{eq:gu}-\ref{eq:shop-attention}. 
The intuition is to find a projection matrix that embeds (via $\mathbf{W}^t \in R^{T*C}$)  the tags into a common latent space as the image features. If the image feature at one position matches (i.e. aligned well with) the embedded tag feature, we assign a larger attention weight for that position to let it contribute more in the final image representation. We adopt inner-product (Equation~\ref{eq:gu}) as the alignment function and the Softmax function (Equation~\ref{eq:shop-attention}) to generate the weights. The output of the subnetwork is a $C$-dimensional feature vector aggregated across all spatial locations according to Equation~\ref{eq:shop-feature}.

\subsection{Attention Modeling for User Images}\label{sec:user-attention}

For user images, i.e. the query images, they are usually taken occasionally using smart phones. Consequently, the focus of the image may not be clear, e.g. when there are multiple (background) objects in the image. In addition, for some kinds of products like clothes, the images are subject to deformations. It is thus important to locate the attention of the image for extracting effective features. However, the attention is not easy to get without any context information (the tags of the query images are usually not available).

Given a query image and a shop image, we infer the spatial attention of the query image using the shop image as the context. Denote $\mathbf{x}\in R^C$ as the output of \Romannum{4}, which could be either a positive image $p$ or a negative image. Following Equation~\ref{eq:user-attention}-\ref{eq:user-feature}, we infer the attention of the anchor image $o$ in subnetwork \Romannum{5}.

\begin{eqnarray}
    g_u(\mathbf{o}_l, \mathbf{x}) &=& \mathbf{v}\cdot \mathbf{o}_l + \mathbf{U}_{l\cdot}\cdot \mathbf{x} \label{eq:gs}\\
	a_l &=& \frac{e^{g_u(\mathbf{o}_l, \mathbf{x})}}{\sum_{k}^L e^{g_u(\mathbf{o}_k, \mathbf{x)}}} \label{eq:user-attention}\\
	f_{\Romannum{5}}(\mathbf{o}, \mathbf{a})&=&\sum_l^L a_l*\mathbf{o}_l \label{eq:user-feature}
\end{eqnarray}
where Equation~\ref{eq:gs} is a linear alignment function, which has better performance than the inner-product alignment function for our experiment. $\mathbf{v} \in R^C$, $\mathbf{U} \in R^{L*C}$ are weights to learn. 

\subsection{Loss Function}

We adapt the triplet loss function as the training objective, which is shown in Equation~\ref{eq:loss}.

\begin{equation}
	L(o, p, q) = \max(0, d(\mathbf{o}^p,\mathbf{p}) - d(\mathbf{o}^q, \mathbf{q}) + \alpha) \label{eq:loss}
\end{equation}
where $d(\cdot,\cdot)$ measures the distance of two images based on their final representation. Euclidean distance is used for $d(\cdot,\cdot)$. Following \cite{schroff2015facenet}, in order to make the training converge stable, we normalize the output of subnetwork \Romannum{4} and \Romannum{5} via L2 norm before feeding them into the loss function. The margin value $\alpha$ is tuned on a validation dataset (0.5 for our experiments). 
Different to the existing approaches~\cite{liu2016deepfashion,huang2015cross} that use the same representation for $o$ in Equation~\ref{eq:loss}, i.e. $d(\mathbf{o,p})$ and $d(\mathbf{o,q})$. We have different representations for $o$, i.e. $\mathbf{o}^p$ and $\mathbf{o}^q$ for $d(\mathbf{o,p})$ and $d(\mathbf{o,q})$ respectively. This is because two sets of attention weights are generated against $p$ and $q$ respectively. The loss function penalizes the triples if $d(\mathbf{o}^p,\mathbf{p}) +\alpha > d(\mathbf{o}^q,\mathbf{q})$ by updating the model parameters to make matched images close and unmatched images far away in the embedded Euclidean space.

\section{Experiments} \label{sec:exp}
In this section, we conduct experimental study by comparing our proposed approaches in Section~\ref{sec:approach} with baseline methods in terms of search effectiveness and efficiency on two real-life datasets.

\subsection{Dataset} \label{sec:data}
In our experimental study, we use the DARN dataset~\cite{huang2015cross} and DeepFashion dataset~\cite{liu2016deepfashion}. 

The DARN dataset is collected for street-to-shop retrieval, i.e. matching street images taken by users with professional photos provided by online shopping sites. After removing corrupted images, we get a subset of 62,812 street images and 238,499 shop images of 13598 distinct products distributed over 20 categories. Each street image has a matched shop image. This dataset also provides semantic attributes for the products. Detailed information of the attributes is available in \cite{huang2015cross}. 
We use 7 types of attributes except the color attribute since we observe that same product may be displayed with different colors in this dataset. 
We partition the dataset into three subsets for training, validation and test, with no overlap of products (see Table~\ref{tb:dataset}). 

The DeepFashion dataset includes over 800,000 images with various labeled information in terms of categories, clothes attributes, landmarks, and image correspondences for cross-domain/in-shop scenario. We do not explore the landmark information since it is beyond the scope of this paper. Additionally, we only use a subset of images from street2shop set, i.e. 19,387 distinct upper clothes of over 130,000 images. We select 11 types of tags that are related to the upper clothes of the product images. 
Similar partition method is applied for this dataset (see Table~\ref{tb:dataset}).
Different from DARN dataset, DeepFashion dataset has more street images (130,000+) than shop images (21,377).

\begin{table}
	\centering
	\caption{Dataset Partition}
	\label{tb:dataset}
	\begin{tabular}{|p{4cm}|c|c|}
		\hline
		Dataset & DARN & DeepFashion \\ \hline
		Distinct Training Products & 10,979 & 15,494 \\ \hline
		Training Street Photos & 50,528 & 84,161 \\ \hline
		Training Shop Photos & 32,194 & 21,377 \\ \hline
		Distinct Validation Products & 9,635 & 1,948 \\ \hline
		Validation Street Photos & 6,318 & 1,948 \\ \hline
		Validation Shop Photos & 23,828 & 2,694 \\ \hline
		Distinct Test Products & 9,636 & 1,945 \\ \hline
		Test Street Photos & 5,966 & 1,945 \\ \hline
		Test Shop Photos & 23,773 & 2,637 \\
		\hline
	\end{tabular}
\end{table}

\subsection{Approaches}\label{sec:baseline}

\subsubsection{Baseline Approaches}
\begin{enumerate}
	\item \textbf{DARN}\cite{huang2015cross}\footnote{We use the notation DARN for both the dataset and the method.}
	has two branches of NIN (Network in Network) networks, one for the street domain and one for the shop domain. It is different to YNet which shares the same bottom layers for both domains. On top of the NIN networks, there are several fully connected layers for category and attribute prediction. The training loss is a weighted combination of the prediction losses and the triplet loss. The triplet loss is calculated using a long feature vector by concatenating the features from convolution layers and fully connected layers.
	We adjust the shapes of lower convolution layers to be the same as the original NIN model~\cite{lin2013network} in order to use the pretrained parameters of NIN over ImageNet. 
	
	\item \textbf{FashionNet}\cite{liu2016deepfashion}
	FashionNet shares the convolution layers for both domains. It has a landmark prediction subnetwork whose results are used to subsample the feature maps of the last convolution layer. The top branches are for different tasks (including tag prediction and landmark prediction) instead of for different domains as in YNet. In other words, all images are passed through the same set of layers in FasionNet, whereas street and shop images are passed through different top layers in YNet. Due to the memory limit, we replace the VGG-16 model~\cite{simonyan2014very} used in the DeepFashion paper with VGG-CNN-S~\cite{chatfield2014return} \footnote{https://gist.github.com/ksimonyan/fd8800eeb36e276cd6f9}. We also remove the landmark prediction subnetwork since exploring the effect of landmark is out of the scope of this paper.
	
	\item \textbf{TripletNIN} and \textbf{TripletVGG} By removing the category and attribute prediction subnetworks of DARN and FashionNet, we get two networks whose loss function only includes triplet loss.  
	\item \textbf{NIN} and \textbf{VGG-CNN-S} We use the NIN and VGG-CNN-S trained on ImageNet to extract the feature for both user images and shop images.
\end{enumerate}


\subsubsection{Network Configuration of Our Approaches}

\begin{enumerate}
	\item \textbf{TagYNIN} and \textbf{CtxYNIN} based on NIN. TagYNIN uses the first 4 convolution layer blocks of NIN as subnetwork \Romannum{1}. The 5-th convolution layer block is used for subnetwork \Romannum{2} and \Romannum{3}. Subnetwork \Romannum{4} uses inner-product as the attention alignment function. The output of subnetwork \Romannum{2} and \Romannum{4} are fed into the triplet loss. CtxYNIN adds the subnetwork \Romannum{5} on top of TagYNIN. 
	\item \textbf{TagYVGG} and \textbf{CtxYVGG} based on VGG-CNN-S. TagYVGG is similar to TagYNIN except the NIN layers are replaced with layers from VGG-CNN-S.
	CtxYVGG is similar to CtxYNIN except the NIN layers are replaced with layers from VGG-CNN-S.
\end{enumerate}

\subsubsection{Implementation Details}

We use backpropagation for parameter gradient calculation and mini-batch Stochastic Gradient Descent (SGD) for parameter value updating. The batch size is 32 and momentum is 0.9.
Learning rate is set to 0.01 at the initial state and decays by 0.1 for every 30 epochs. Margin in the loss function is set to 0.5 for the baseline experiments. We set the margin to 0.3 for YNet and TagYNet, and 0.5 for CtxYNet. We train our networks in the order of YNet, TagYNet, and CtxYNet by using the parameters trained from the previous network to initialize the subsequent network. The pretrained parameters over ImageNet (e.g. using NIN) is used to initialize the corresponding YNet (e.g. YNIN). For YNet and TagYNet, we use the standard triplet loss function; For CtxYNet, we use the adapted triplet loss function (Equation~\ref{eq:loss}). When training TagYNet and CtxYNet, we freeze the parameters in subnetwork \Romannum{1}. Considering that bounding boxes and landmarks are costly to annotate for large image databases in real applications, we do not use these information for training. In contrast, the attribute and category information is typically available on e-commerce websites. Therefore, we use them in training DARN, FashionNet and our approaches.

\subsection{Evaluation Metric}
Following \cite{huang2015cross,liu2016deepfashion}, we evaluate the retrieval performance by top-K precision, which is defined as follows: 
\begin{equation}
P@K = \frac{\sum_{q\in Q}hit(q, K)}{|Q|}
\end{equation}
where $Q$ is the total number of queries; $hit(q, K)=1$ if at least one image of the same product as the query image $q$ appears in the returned top-K ranking list; otherwise $hit(q, K)=0$. For most queries, there is only one matched database image in both the DARN and DeepFashion datasets.


\subsection{Comparison on DARN Dataset}
\begin{figure}
	\centering
	\includegraphics[width=0.4\textwidth]{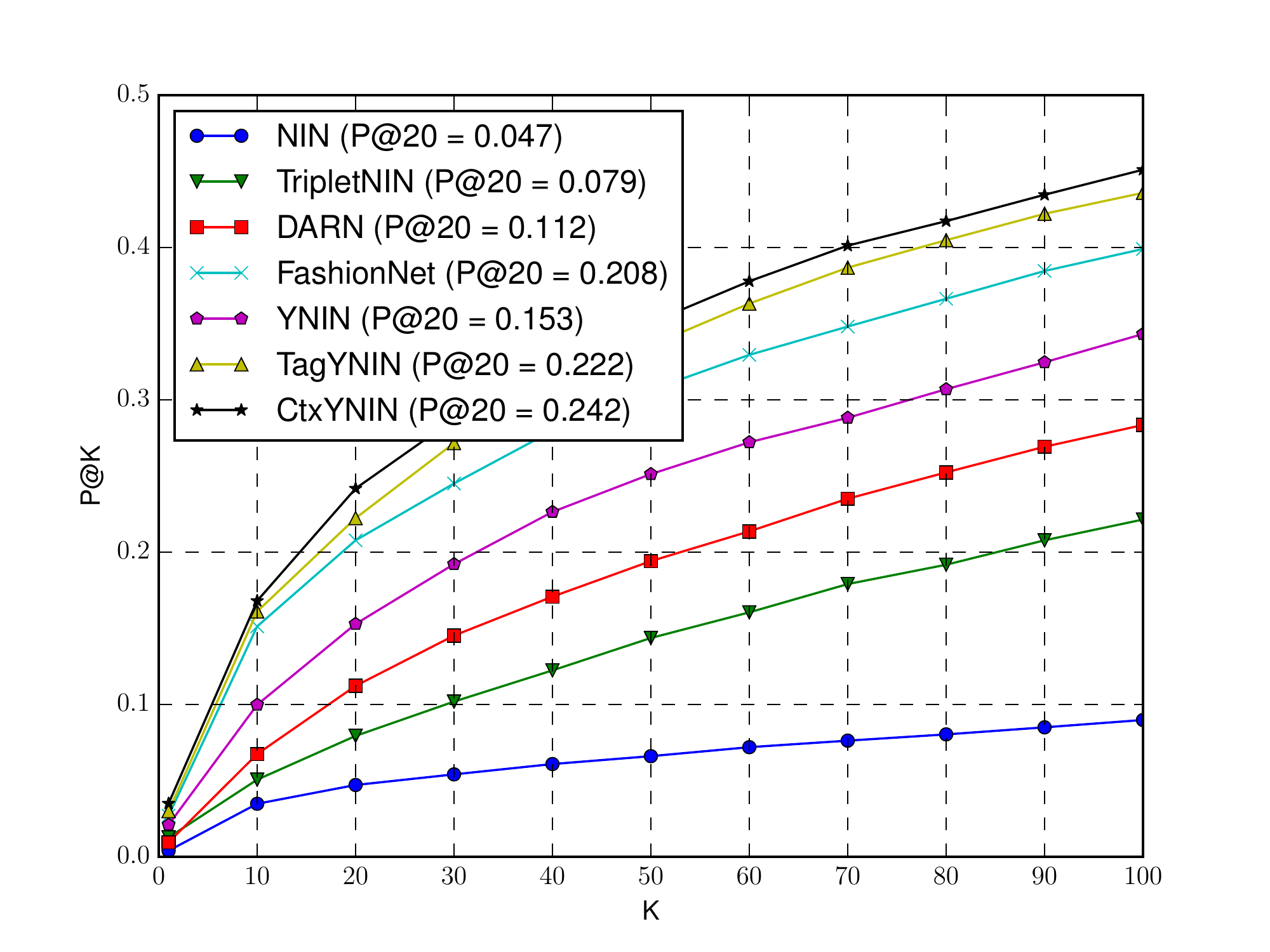}
	\caption{Comparison of P@K on DARN dataset.}\label{fig:darn}
\end{figure}

Figure~\ref{fig:darn} shows the top-K precision of the baseline approaches and our approaches on DARN dataset. We can see that the pretrained NIN performs worst since the task and dataset are different to our product image retrieval application.
TripletNIN is better than NIN as it is fine-tuned with metric learning over the DARN dataset to distinguish the inner-category images (e.g. different T-shirts). However, it is not as good as DARN which considers the tag information during training through attribute and category prediction. Meanwhile, we can observe that our TagNIN and CtxNIN outperform a lot than YNIN which does not incorporate tag information. These prove the effectiveness of exploring tag information for image retrieval.

We further compare DARN, FashionNet and our attention modeling based approaches. From Figure~\ref{fig:darn}, we observe that both of our attention modeling approaches significantly outperform DARN (over 10\% improvement on top-20 accuracy) and FashionNet (1.4\% and 3.4\% improvement on top-20 accuracy). 
DARN and FashionNet only use tags during training, which serves as a regularization on the extracted feature (require the feature to capture tag information). Our TagYNet uses tags in both training and querying phases for the shop images, which helps to locate the attention of the shop images especially when noisy background or multiple products occur in one image. Consequently, it captures more discriminative features. In addition, we do not need to tune the weights of prediction loss and triplet loss as in DARN and FashionNet. In terms of network architecture, our YNet architecture involves fewer parameters compared with the dual network architecture of DARN, and thus is more robust to overfitting. 

CtxYNet performs slightly better than TagYNet by reranking the top-256 list retrieved from TagYNet.
It indicates that attention modeling of the query images indeed helps to obtain better feature representations. 



\subsection{Comparison on DeepFashion Dataset}
\begin{figure}
	\centering
	\includegraphics[width=0.4\textwidth]{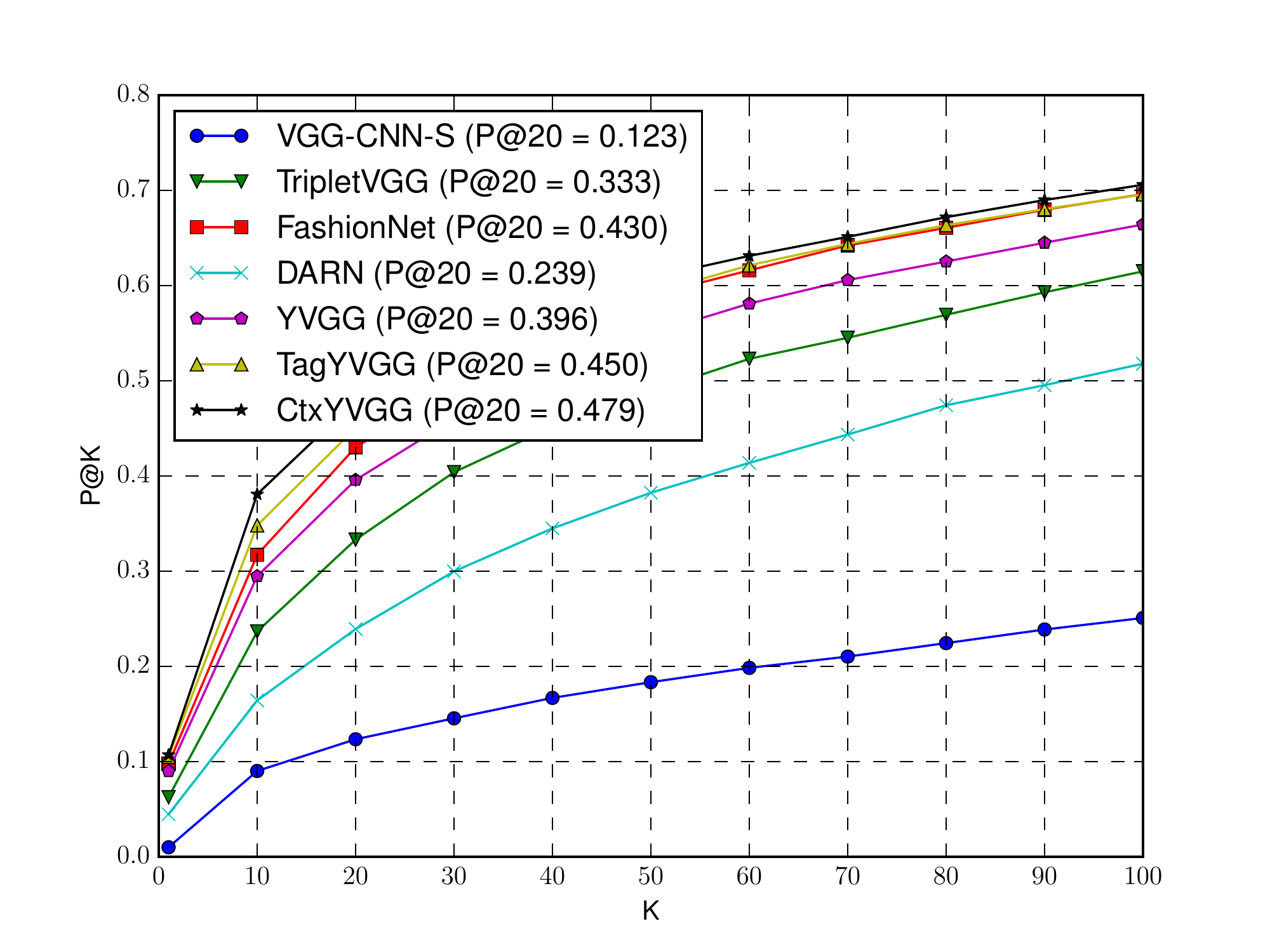}
	\caption{Comparison of P@K on DeepFashion dataset.}\label{fig:deepfashion}
\end{figure}

Figure \ref{fig:deepfashion} shows the detailed top-k precision of baseline approaches and our approaches on DeepFashion dataset. 
We notice that the retrieval performance of all approaches on this dataset is much better than those on DARN's dataset, mainly due to the smaller size of the database. 

We observe that our attention modeling approaches again perform best over the other baselines approaches. Compared with DARN, FashionNet achieves significantly better performance than DARN, which is consistent with results shown in the paper \cite{liu2016deepfashion}. However, our proposed network performs even better than FashionNet. Our tag-based attention mechanism (TagYNet) and context-based attention mechanism (CtxYnet) gain 2\% and 4\% improvement on top-20 precision over FashionNet.
There are still some noisy tags (not visually relevant to the image content, e.g. 'thickness of clothes') in DeepFashion dataset, therefore, we expect to see better performance if these tags/attributes are filtered out.

\subsection{Query Efficiency}
\begin{figure}
	\centering
	\includegraphics[width=0.36\textwidth]{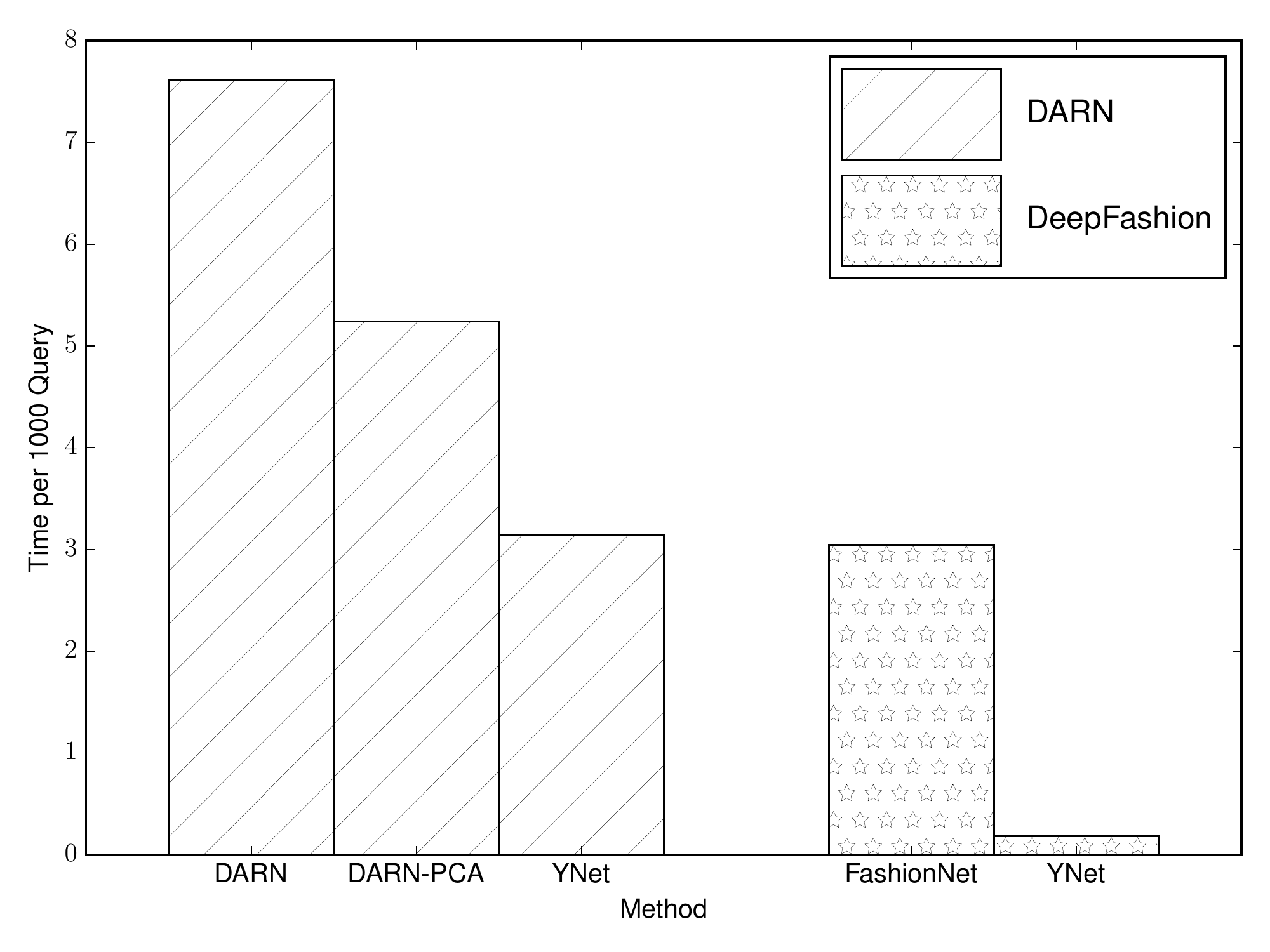}
	\caption{Querying efficiency with different feature dimensions and different size of databases.}\label{fig:effi}
\end{figure}

\begin{figure}
	\begin{subfigure}{.5\textwidth}
		\centering
		\includegraphics[width=.9\linewidth]{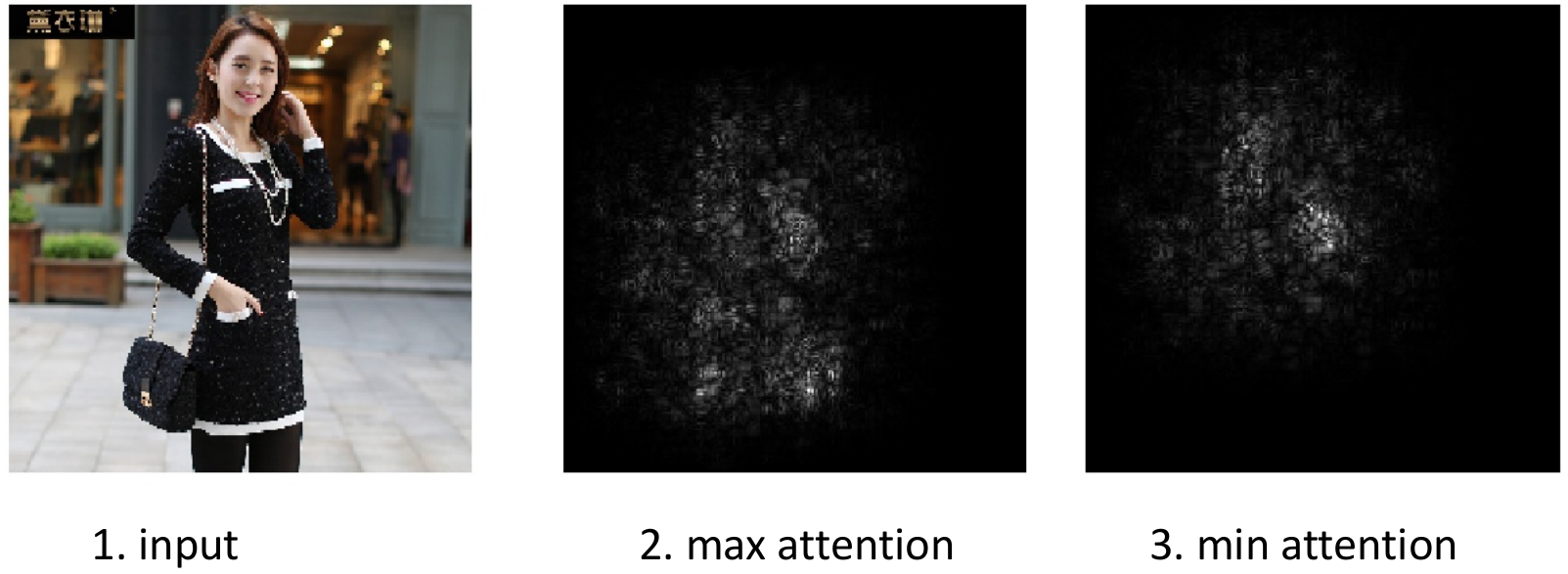}
		\caption{\ }
		\label{fig:vis1}
	\end{subfigure}%
	\\
	\begin{subfigure}{.5\textwidth}
		\centering
		\includegraphics[width=.9\linewidth]{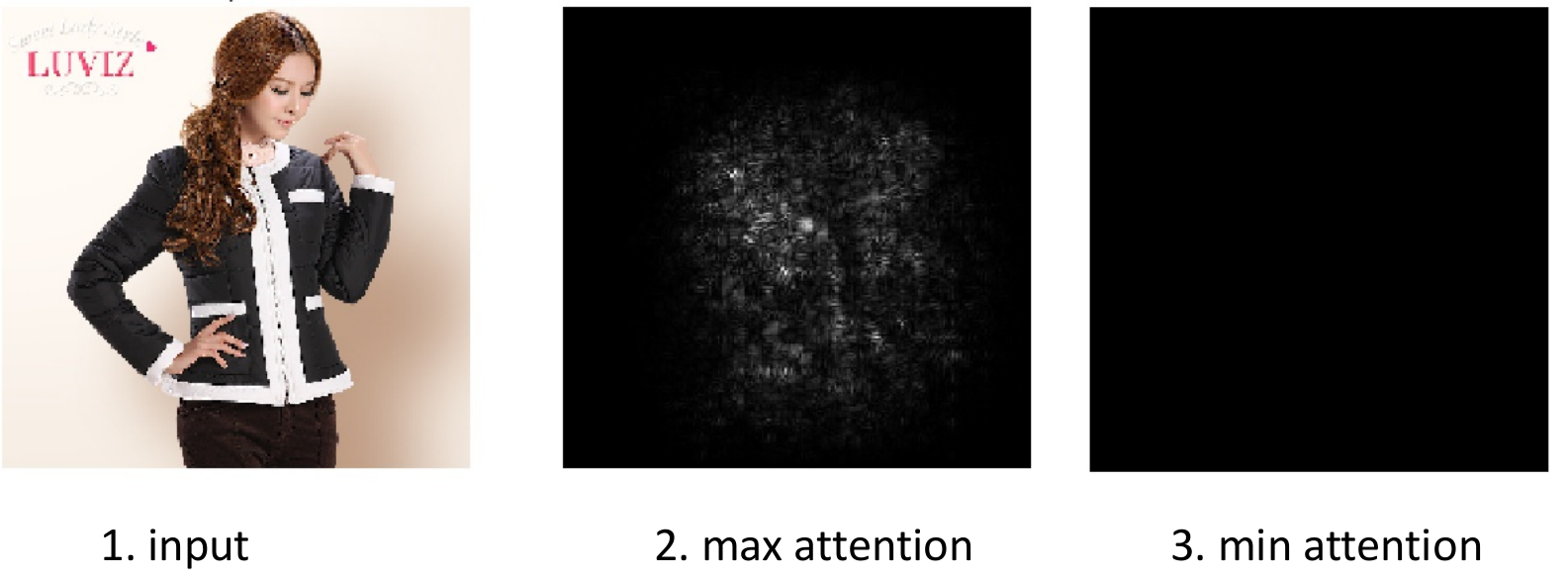}
		\caption{\ }
		\label{fig:vis2}
	\end{subfigure}
	\caption{Images and their attention maps.}
	\label{fig:vis}
\end{figure}

\begin{figure*}
	\centering
	\includegraphics[width=0.7\textwidth]{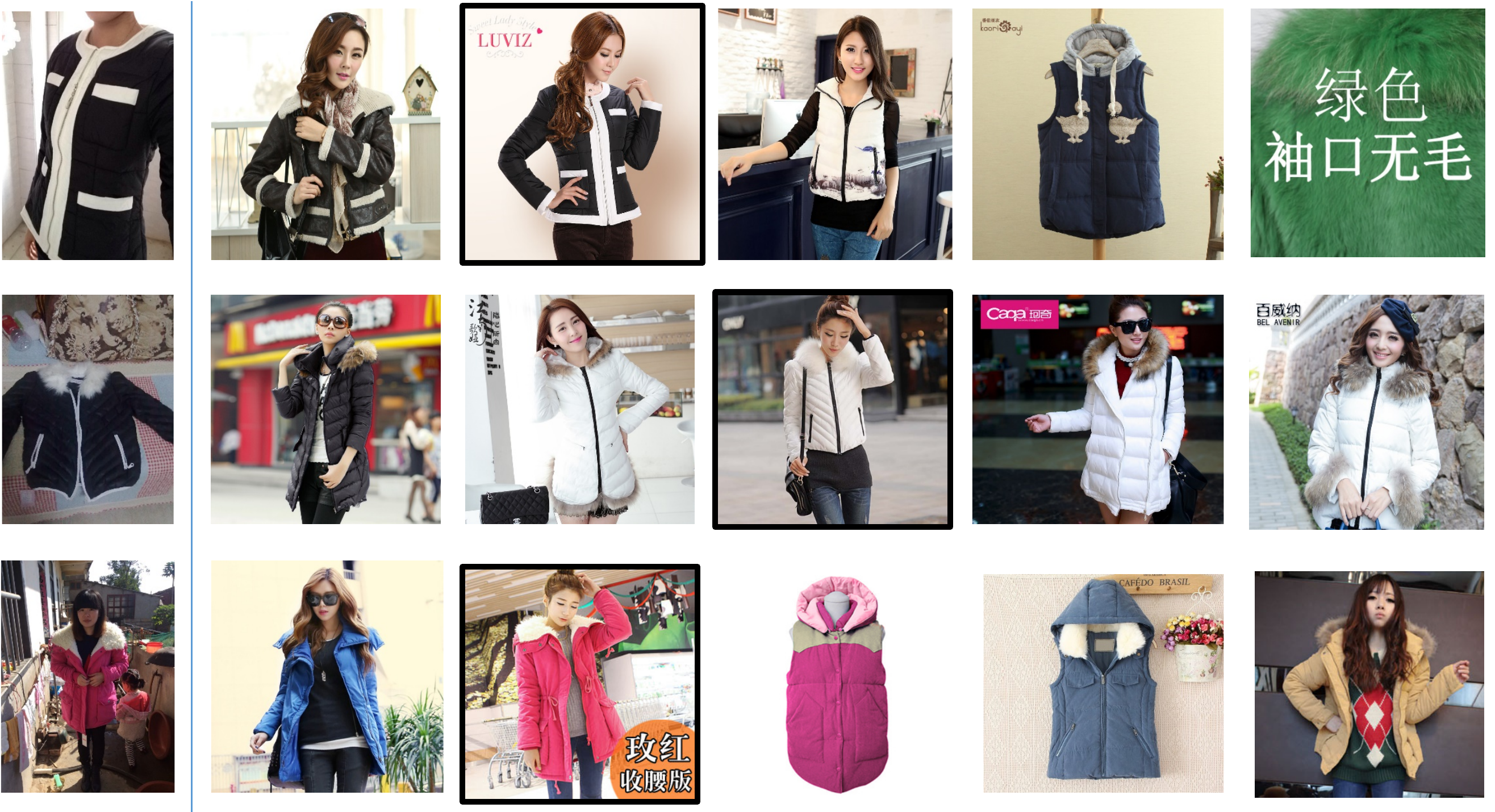}
	\caption{Example queries (the first column) with matched results (with border) in top-5 list.}\label{fig:good}
\end{figure*}

\begin{figure*}
	\centering
	\includegraphics[width=0.7\textwidth]{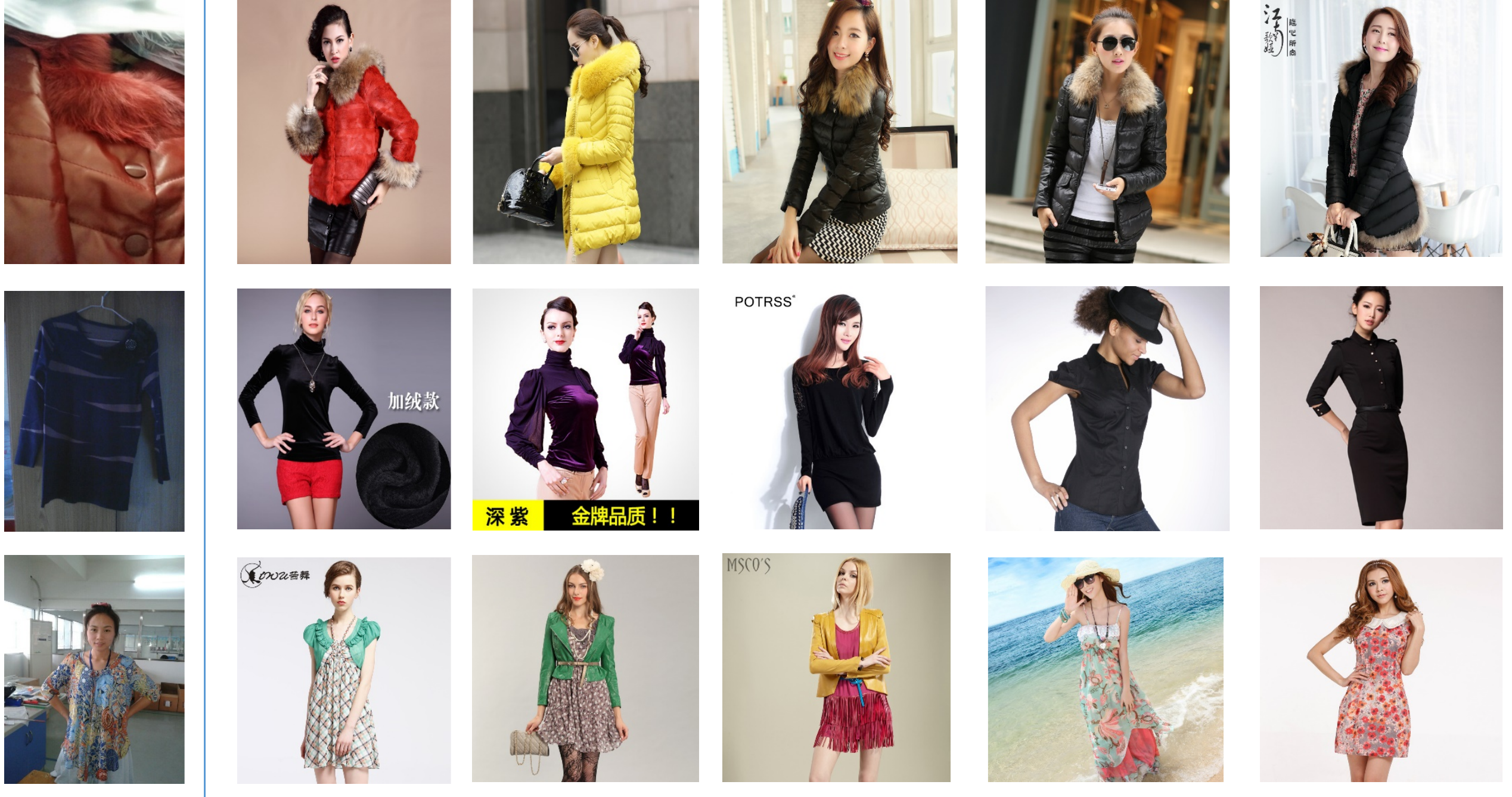}
	\caption{Example queries (the first column) without matched results in top-5 list.}\label{fig:bad}
\end{figure*}

Our retrieval system runs on a server with Intel i7-4930K CPU and three GTX Titan X GPU cards. We develop the system using SINGA~\cite{ooi2015singa,wang2016deep} --- a deep learning library.

We measure the query efficiency on the two datasets as shown in Figure \ref{fig:effi}. The feature extracted by our approaches is aggregated from the features across spatial locations of the convolution layer. Therefore, the feature dimension is the number of the feature maps in the last convolution layer, which is 1000 and 512 for NIN-based and VGG-based networks respectively. DARN and FashionNet\cite{huang2015cross,liu2016deep} concatenate the local features from convolution layers and global features from fully connected layers. Consequently, their feature dimension is larger than ours. Even after applying PCA, the feature dimension is still up to 8196-D, which is 8x and 16x larger than our NIN-based and VGG-based features, respectively. As a result, our approaches achieve better efficiency as shown in Figure~\ref{fig:effi}.

\subsection{Attention Visualization}

For better understanding of the attention modeling mechanism, we visualize the attention maps of two sample images from DARN in Figure~\ref{fig:vis}. The `max attention' (resp. min attention) map is generated following \cite{simonyan2013deep}: first, setting the gradients of the final convolution layer as 0 except those for the location (denoted as $l$) with the maximum (resp. minimum) attention weight computed from TagNIN, whose gradients are set to 1; second, back-propagating the gradients to get the gradients of the input data, which reflect the activations of location $l$ and are used to plot the attention map. From Figure~\ref{fig:vis1}, we can see that the activations of the maximum weighted position matches the attention of the input image better than that of the minimum weighted position. In addition, the noisy background is filtered in the attention maps, where the activations mainly cover the clothes. In other words, the attention weights have semantic explanations.

\subsection{Example Queries and Results}

We analyze some sample queries and corresponding results for better understanding the task and feature extraction models. The DARN dataset and TagYNet are used for this experiment. Four queries with matched results in the top-5 list are shown In Figure\ref{fig:good}. First, we can see that the result images (i.e. shop images in column 2-6) have much better quality than the query images (i.e. user image in the first column). Therefore, we need separate branches for extracting the features from the two domains. Second, some shop images, e.g. the matched result of the third query, also have noisy background. It is necessary to incorporate extra information like tags to locate the image attention. Third, the query image and the matched result may look quite different from the global view, as shown in the second row. In spite of such difference, our model is able to focus on local pattens (e.g. the collar and pocket) to find matched results. 

We also sample some queries for which our model fails to find matched images in the top-5 list, as shown in Figure~\ref{fig:bad}. First, we can see that these query images are either cropped (the first query) or taken under insufficient lighting (the second query). Second, the evaluation criteria is very strict. As shown in the third row, some images are not considered as matched results although they are visually very similar to the query images. In fact, only images of the exact same product as the query are considered matched results.

\section{Conclusion}
To tackle the problem of cross-domain product image search, we have presented a novel neural network architecture which shares bottom convolutional layers to learn domain-invariant features and separates top convolutional layers to learn domain-specific features. Different from other approaches, we introduce a tag-based attention modeling mechanism denoted as TagYNet by exploiting product tag information, and a context-based attention modeling mechanism denoted as CtxYNet, using the candidate image as the context. Experiments on two public datasets confirm the efficiency and effectiveness of the features extracted using our attention based approaches. 

\section{Acknowledgment}
This work is supported by the National Research Foundation, Prime Minister's Office, Singapore, under its Competitive Research Programme (CRP Award No. NRF-CRP8-2011-08), and FY2017 SUG Grant, National University of Singapore.

\bibliographystyle{ACM-Reference-Format}
\balance
\bibliography{ref} 

\end{document}